\documentstyle[aps,prl,epsf,floats]{revtex} 
\draft

\newcommand{\beq}{\begin{equation}}
\newcommand{\eeq}{\end{equation}}
\newcommand{\bea}{\begin{eqnarray}}
\newcommand{\eea}{\end{eqnarray}}

\newcommand{\lsim}{\stackrel{<}{\scriptstyle \sim}}
\newcommand{\gsim}{\stackrel{>}{\scriptstyle \sim}}

\begin{document}
\twocolumn[\hsize\textwidth\columnwidth\hsize\csname
@twocolumnfalse\endcsname 


\preprint{UCLA/01/TEP/7; 
} 

\title{Experimental identification of non-pointlike dark-matter candidates 
} 

\author{Graciela Gelmini$^1$, Alexander Kusenko$^{1,2}$, and Shmuel
Nussinov$^3$ } 

\address{$^1$Department of Physics and Astronomy, UCLA, Los
Angeles, CA 90095-1547 \\ $^2$RIKEN BNL Research Center, Brookhaven
National Laboratory, Upton, NY 11973 \\ $^3$Sackler Faculty of Science, Tel
Aviv University, Tel Aviv, Israel }

\date{March, 2002}

\maketitle
             
\begin{abstract}

We show that direct dark matter detection experiments can distinguish
between pointlike and non-pointlike dark-matter candidates.  The shape of
the nuclear recoil energy spectrum from pointlike dark-matter particles,
{\em e.\,g.}, neutralinos, is determined by the velocity distribution of
dark matter in the galactic halo and by nuclear form factors.  In contrast,
typical cross sections of non-pointlike dark matter, for example, Q-balls,
have a new form factor, which decreases rapidly with the recoil energy.
Therefore, a signal from non-pointlike dark matter is expected to peak near
the experimental threshold and to fall off rapidly at higher 
energies.  Although the width of the signal is practically independent of
the dark matter velocity dispersion, its height is expected to exhibit an
annual modulation due to the changes in the dark-matter flux.  

\end{abstract}

\pacs{PACS numbers: 95.35.+d, 98.80.Cq   \hspace{1.0cm} UCLA/01/TEP/7} 

\vskip2.0pc]

\renewcommand{\thefootnote}{\arabic{footnote}}
\setcounter{footnote}{0}

Direct detection of weakly interacting dark-matter particles is the goal of
several ongoing and planned experiments, which try to observe the nuclear
recoils from interactions with dark matter.  In this letter we will show
that, in addition to measuring the cross section and the flux of
dark-matter particles, the same experiments may be able to probe the
structure of dark matter and distinguish between pointlike and extended
objects, both of which have been proposed as dark-matter candidates.

A number of reasons have led one to consider pointlike and non-pointlike
candidates for dark matter.  For example, supersymmetric extensions of the
Standard Model predict both the pointlike dark matter in the form of the
lightest supersymmetric particles (LSP)~\cite{jgk}, as well as a
non-pointlike dark matter in the form of SUSY Q-balls~\cite{ak_mssm,ks}.
The population of LSP could be produced by their freeze-out from
equilibrium, while Q-balls could be copiously generated by fragmentation of
the Affleck-Dine condensate~\cite{ks}.  Mirror atoms have also been
proposed as extended dark matter candidates~\cite{mirror}.  Recent analyses
of dark matter halos have provided an additional motivation for considering
extended objects as dark matter~\cite{sidm}.  Numerical simulations with
improved resolution seem to predict an excessive number of subhalos and
satellite galaxies~\cite{MooreSub}, as well as a cuspy distribution of dark
matter, which is at variance with some observations~\cite{sidm}.  However,
recent measurements of substructure in gravitational lens galaxies show
agreement with N-body simulations~\cite{dalal}. 
Self-interacting dark matter~\cite{sidm,sidm_examples,ak_ps} could
erase small-scale structure.  
However, the required cross section
for self-interaction is so large that, for pointlike particles, it violates
unitarity, unless they have masses of the order of a few GeV or
lighter~\cite{unitarity}.

This limit does not apply, however, to the extended objects, such as
Q-balls, which could be self-interacting dark matter with a wide range of
masses $M_{_D}$ and sizes $R$~\cite{ak_ps}.  Their cross section of
self-interaction can be as large as their geometrical size,
$\sigma_{{_D}{_D}}\simeq R^2$, and can be in the range necessary for removing
the cusp and the galactic satellites~\cite{sidm}:
\beq
\sigma_{{_D}{_D}}= (0.8-10)\times 10^{-24} (M_{_D}/{\rm GeV}) {\rm cm}^2.
\label{sDD} 
\eeq  
Mirror atoms~\cite{mirror} are also not a subject to the unitarity bound.  

Various dark matter candidates, pointlike or not, can have interactions
with ordinary matter.  Although there are several limits on {\em strong}
matter-dark-matter interactions~\cite{ak_ps}, those interactions that do
not exceed the strength of standard weak interactions are definitely
allowed.  Therefore, experiments designed to detect weakly interacting
massive particles (WIMP) can detect both pointlike particles and extended
objects.

The challenge of identifying the dark matter in the universe requires, in
particular, an answer to the question of whether the dark matter particles
are pointlike or not.  We will show that, for a certain range of masses
and sizes of dark-matter particles, this question can be answered based on
the shape of the nuclear recoil energy spectrum of the signal in direct 
detection experiments. 

The very nature of extended objects implies that the cross section of their
interactions with matter includes a form factor whose dependence on the
momentum transfer is an identifiable signature of a non-pointlike object.
Although the interaction strength is unknown, if such interactions are
detected, the extended objects can be identified by a signal with a recoil
energy spectrum strongly peaked at low energies.  Thus, a WIMP experiment 
should observe a signal that is strong near the threshold and falls off so 
rapidly at higher energies that, in a realistic detector, only one energy
bin should show an annual modulation. 

Although this behavior of the cross section is rather general, we will
discuss Q-balls, for definiteness.  Let us assume that the dark matter
particles are Q-balls, non-topological solitons made of a scalar field
$\phi$.  A Q-ball is a coherent state of $Q$ quanta of the $\phi$ field.
We denote its mass $M_{_D}=M_{_D}(Q)$.  

For simplicity, let us assume that the dark matter particles scatter off
nuclei via the exchange of some heavy boson $Z'$, the analog of $Z$
exchange in neutralino-nuclear scattering.  While the resulting interaction
is almost pointlike at some fundamental level, the
extended structure of the DM particle causes a form factor $F_D$ to
appear in the elastic scattering amplitude.  This form factor is analogous
to the nuclear form factor $F_N(A)$ that reflects the nuclear structure.

The direct detection strategy for WIMPs relies on detecting the recoil of
nuclei with atomic number A and mass $M_{_N}=M_{_N}(A)$.  Dark-matter
particles have velocities $\beta = v/c \simeq 10^{-3} $, so the maximum
energy deposited in a collision is 
\beq 
\Delta E =2 \frac{ \beta^2 \mu^2} {M_{_N}(A)},
\eeq 
where $\mu$ is the reduced mass of the dark matter particle, with mass
$M_{_D}(Q)$, and the target nucleus, with mass $M_{_N}(A)$. Existing and
future detectors, have energy thresholds of a few keV.  Hence, we require
that $\Delta E \gsim 10$~keV.  For $M_{_N}(A) \simeq 100$~GeV, this implies
$M_{_D}(Q) \gsim M_{_N}(A)$. We, therefore, restrict our discussion to
Q-balls heavier than nuclei, for which
\beq 
\Delta E \simeq 2 \beta^2 M_{_N}(A) .
\eeq 
The differential cross section for elastic scattering is
\beq
\frac{d \sigma}{d q^2}= \frac{(G'_{_F})^2}{4\pi}
\frac{M_{_{Z'}}^4}{\left (q^2-M_{_{Z'}}^2 \right)^2} \frac{A^2}{\beta^2}
Q^2 \ |F_A|^2 \ |F_D|^2, 
\label{dsdq2} 
\eeq
where, $q$ is the momentum transfer, $G'_{_F}$ is the analogue of the Fermi 
constant for $Z'$ exchange,
while $F_A$ and $F_D$ are the form factors of the nucleus and the Q-ball,
respectively:
\beq
F_{D,A}(q) = \int d^3 \vec{r} \ e^{i\vec{q} \vec{r}} \ 
\rho_{{_D},{_A}}(\vec{r}).  
\label{FF} 
\eeq
To a good approximation, the form factor of a nucleus can be calculated
using $\rho_{_A}= \rho_0/\left (1+ \exp\{ (r-c)/z \} \right ) $, where, for
example, for Ge, $c=4.503, z=0.583$.

If (i) the size of a DM particle exceeds the size of the nucleus, and (ii)
the putative interaction with quarks allows a coherent channel, then the
form factor $F_D$ imprints the cross section (\ref{dsdq2}) with an easily
identifiable signature: a rapid fall-off for large $q$.

For definiteness we assume that the cross section $\sigma_{{_D}{_D}}$ and
mass $M_{_D}$ satisfy eq.~(\ref{sDD}), which relates the Q-ball
mass and its radius:  
\beq
R\simeq (1-3)\times 10^{-12} {\rm cm} \left ( \frac{M_{_D}}{\rm GeV} \right
)^{1/2}. 
\label{RM}
\eeq

We now demand that the collision be elastic.  Q-balls have nearly massless
modes, for example, the Goldstone mode of a spontaneously broken U(1)
symmetry. The lowest excitation has a Compton wavelength of the order $R$
and energy gap $\simeq 1/R$.  The collision is elastic as long as the energy
transfer, which is at most $2 \beta^2 M_{_N}(A)$, is smaller than
$1/R$. Hence, the collision is always elastic for 
\beq 
R < \frac{1}{2 \beta^2 M_{_N}(A)} \simeq \frac{1}{0.1 {\rm MeV}} \simeq
1 \times 10^{-10} {\rm cm} 
\eeq
Using eq.~(\ref{RM}), we get the upper limit on the mass of Q-balls that
scatter elastically, $M_{_D}<10^4$~GeV.  As mentioned above, a detection
threshold of a few keV requires the Q-ball to have a mass similar or 
larger than the nucleous mass, which is of the order of $10^2$~GeV.

Hence, we restrict our discussion to the range\footnote{
In the case of mirror matter, if the $Z'$ is coupled to both the 
ordinary quarks and the electrons, the ordinary atoms would scatter
coherently off a mirror atom.  In this case the range of radii for which
the scattering is elastic and coherent extends
as far as  to $10^{-11}{\rm cm} \lsim   R  \lsim 10^{-8}{\rm cm}$. 
} 
\bea
10^2 {\rm GeV} \lsim & M_{_D} & \lsim 10^4 {\rm GeV}, \label{rangeM} \\
10^{-11}{\rm cm} 
\lsim  & R & \lsim 10^{-10}{\rm cm} 
\label{rangeR}
\eea

The form factor $F_D$ in eq.~(\ref{FF}) can be evaluated for a given Q-ball
profile.  Depending on the scalar potential $U(\phi)$, Q-balls can have a
thin-wall~\cite{Qthin} or a thick-wall profile~\cite{Qthick}.  However, in
either case, for $qR>1$
\beq
F(q) \propto \frac{1}{(qR)^n}, ~~n\ge 2. 
\eeq
$F(q)$ decreases rapidly for $qR>1$.  The slowest possible fall-off
occurs when the Q-ball profile is approximated by a step-function.  In this
case,
\beq
F_{\rm step} (q) \propto \left (\frac{qR \cos(qR) - \sin(qR)}{(qR)^3}
\right)
\eeq
which fall off as $(qR)^{-2}$,  for $qR>1$. 

For realistic profiles, $F(q)$ need not have the oscillating behavior 
of eq.~(11) and,
more importantly, it has a faster asymptotic fall-off, {\em e.\,g},
$\sim(qR)^{-3}$ or faster.  

The key observation is that, for the range of parameters in
eqs.~(\ref{rangeM}-\ref{rangeR}), $(qR)>1$ in typical direct detection
experiments.  Indeed, the maximum momentum transfer in an elastic collision
is $q_{max}\simeq 2 \beta M_{_A} \simeq 200$~MeV.  On the other hand, the
minimal observable energy of recoil is determined by the experimental
threshold of a few keV.  For $M_D > M_A$, this means $q^2/ 2 M_A > 1$~keV,
or $q \gsim 10$ MeV. Hence, from eqs.~(\ref{rangeM}-\ref{rangeR}), 
\beq
5 \lsim (qR) \lsim 10^3.
\label{rangeqR}
\eeq 
The $q$-dependence of the differential cross section (\ref{dsdq2}) is,
therefore, dominated by $|F_D|^2 \sim (qR)^{-2n}, \, n\ge 2$: 
\beq
\frac{d \sigma}{d q^2} \propto  \frac{1}{(qR)^l}, \ l\ge 4  
\eeq 
Thus, the nuclear recoil energy spectrum falls off very fast with
increasing energy, independent of the incident DM velocity
distribution. The signal is dominated by the low-energy events near the
threshold. The magnitude of this signal should be modulated by the annual
variation of the DM velocity distribution.  In contrast, the signal from
pointlike dark matter would have a wider recoil-energy profile determined
by the velocity distribution of dark matter particles and the nuclear form
factor.  Halo models with effectively maxwellian distribution, including
those with triaxiality, velocity anisotropy, or clumping, predict similar
differential event rates~\cite{Green,frieman}.  For pointlike dark matter
particles with masses above $\sim 20$~GeV, the falloff of the differential
spectrum is much less steep than $1/q^6$.

The reason why the nuclear $F_A$ form factor does not have a similar effect
on the shape of the signal from pointlike dark matter is because the
corresponding value of $(qR)$ is  smaller.  It is essential that a condition
$(qR)>1$ is satisfied for extended objects we are considering.

Finally, we show that, for some reasonable model parameters, one can get
the detection rates for dark-matter Q-balls as high as those for
neutralinos.  The cross section in eq.~(\ref{dsdq2}) is, of course,
model-dependent.  The constant $G_{_F}'$ reflects the strength of some
interactions beyond the Standard Model.  The bounds on new heavy $Z'$
bosons imply that $M(Z') \gsim 500$~GeV.  Let us take $G_{_F}'\simeq
10^{-4}G_{_F} $, which corresponds to a TeV scale of new physics.  The form
factor $|F_D|$ can be taken of the order of $10^{-6}$ for $n=3$ and $(q
R)\simeq 10$, or, equivalently, $n=2$, $(q R)\simeq 30$.  For the cross section
to be of the order of weak cross section, the enhancement from coherent
scattering $Q^2$ must compensate for the smaller coupling and the
form-factor suppression, that is $10^{-4} \times 10^{-6} \times Q^2 \simeq
1$. This is possible for $Q\simeq 10^5$, which is within the range of values
considered in Ref.~\cite{ak_ps}. Using the parameterization from
Ref.~\cite{ak_ps}, Q-ball with mass $M_{_D} \simeq \mu Q\simeq 10^4$~GeV and
radius $R_{_Q} \simeq \mu^{-1} Q^{1/3}\simeq 10^{-10}$cm is within the range of
eqs.~(\ref{rangeM}-\ref{rangeR}) for $Q\simeq 10^5$, $\mu\simeq 0.1$~GeV.

To conclude, direct detection experiments designed to search for weakly
interacting massive particles may be able to discern the spatial extent of
a dark-matter particle in a certain range of parameters. The signal from
extended objects would pile up near the experimental threshold, unlike a
typical signal from a pointlike dark matter. This offers an exciting
possibility to distinguish an extended object from a pointlike particle.
 
This work was supported by in part by U.S. Department of Energy
grant DE-FG03-91ER40662 (GG and AK) and by Israeli National Science
Foundation Grant number 561/99 (SN). SN thanks UCLA for hospitality.

\end{document}